\documentstyle[aps,prl,multicol,epsf]{revtex}

\begin{document}

\title{Scale-free energy dissipation and dynamic phase transition
in stochastic sandpiles}

\author{Bosiljka Tadi\'c$^{\star}$}

\address{
Jo\v{z}ef Stefan Institute,
P.O. Box 3000, 1001 Ljubljana, Slovenia}


\maketitle
\begin{abstract}
\newline

We study numerically scaling properties of the  distribution of cumulative
 energy dissipated in an avalanche and the dynamic phase transition in   a
stochastic directed cellular automaton [B. Tadi\'c and D. Dhar, Phys.
 Rev. Lett. {\bf 79}, 1519 (1997)] in $d=1+1$ dimensions.
In the critical steady state occurring for the  probability of toppling
 $p\ge p^\star$=0.70548, the dissipated energy distribution exhibits
scaling  behavior with new scaling exponents $\tau _E $ and
$D_E$ for slope and cut-off energy, respectively, indicating that the
sandpile surface is a fractal. In contrast to avalanche
 exponents, the energy exponents appear to be $p$-dependent
 in the region $p^\star \le p <1$, however the product $(\tau _E-1)D_E$
remains universal.
We estimate the roughness exponent of the transverse section of the pile
as $\chi =0.44\pm 0.04$.
Critical exponents characterizing the dynamic  phase transition at
$p^\star $ are obtained  by direct simulation and scaling analysis of
the survival probability distribution
and the  average outflow current. The transition belongs to a new universality
class with the critical exponents $\nu_\| =\gamma =1.22 \pm 0.02$,
$\beta =0.56\pm 0.02$  and  $\nu _\bot = 0.761\pm 0.029$, with apparent
violation of  hyperscaling. Generalized  hyperscaling relation
leads to $\beta + \beta ^\prime = (d-1)\nu _\bot $, where $\beta ^\prime
= 0.195 \pm 0.012$ is the exponent governed by the ultimate survival
probability.
\end{abstract}

\begin{multicols}{2}

\section{Introduction}

Open driven systems exhibiting self-organized critical states \cite{Bak_book}
are usually modeled by sandpile-type automata, in which sand grains are added
slowly to the system and its evolution is monitored in terms of collective
sandslides (avalanches). With additional complexity due to stochastic
character of the relaxation rules \cite{stochastic_ca,Al_tracer,LTU,BT_gran},
 stochastic cellular automata models have proven useful in
understanding  certain aspects of granular flow
\cite{Al_tracer,LTU,BT_gran} in realistic granular materials and  in
stochastic biological processes \cite{biol_stochastic}.
On the other hand, sandpile automata models are interesting from the
 theoretical point of view since both role of the dynamic conservation
law---conservation of the number of grains---and emergent spatial
structures can  be easily  monitored.

Recently  a model with probabilistic toppling and directed mass flow
has been proposed \cite{TD}, in which the
probability of toppling $p$ represents a control parameter
originating either from random  variations of sticking
properties between grains, or from stochastic external conditions
(wetting and drying properties). Another realization is related to
stochastic processes in biological systems such as random dispersion of
 particles, which are added from the outside and evacuated
from the system only when its response  lasts longer than a {\it fixed}
time $T_0$, measured on the internal time scale of the process.
 Particles are transferred among communicating cells according to
 probabilistic rules, however, for  response times shorter than $T_0$
they are held inside the system. Therefore, each cell  contains a certain
number of particles, which varies  with time.  The probabilistic character
of the particle transfer between connected cells  can be attributed
to mechanisms  which depend on general condition of the system.
We consider the case $T_0\equiv L$, where $L$  is the linear system size.

The relaxation rules of the model in $d=1+1$ dimensions are:
if $h(i,j)\ge h_c=2$ then {\it with the probability} $p$ the relaxation
occurs as follows:
\begin{equation}
h(i,j)\to h(i,j)-2; h(i+1,j_\pm)\to h(i+1,j_\pm)+1 .
\label{rules}
\end{equation}
Here $h(i,j)$ is the dynamic variable, i.e., the height (number of particles)
 at site $(i,j)$, and $(i+1,j_\pm)$ are neighboring downstream sites
on a 2-dimensional square lattice oriented downwards.  Due to directed
mass transport the dynamics of this model is anisotropic leading to
self-affine relaxation clusters in $d=2$ (see Fig.\ 1).
Thus the model can  also be viewed as directional lines in $d=1+1$
dimensions, in which  the
instability can propagate in one spatial dimension back and forth, whereas
the temporal dimension is strictly directed. The system is driven by adding
particles from the outside, one at a time at a random position along the
first (top) row. Only sites which are connected to toppled sites at
the previous time step are considered as candidates for toppling.
Perturbation can then be transferred from an active (toppled) site to two
 forward neighbors.  Periodic transverse boundaries are assumed
and all candidate sites are updated in parallel.
The probabilistic character of the relaxation rules produces a ragged
structure of heights. [A transverse section through the pile is shown in
Fig.\ 1 (top)].
 Two top particles  from the surface $h(i,j)$
are taken away from the system when the {\it active} site is on the
lower boundary, i.e., $i=L$.

It should be stressed that according to the above relaxation rules
transport of grains is independent of the relative heights of neighboring
sites.
Therefore, it may occur that at some sites height difference in the
direction of transport is negative, and thus the system performs  work
in order to maintain the transport.
A nice example of this dynamic rules was found recently in biological
transport processes which are mediated by
so called molecular motors \cite{biol-motors}. These are protein molecules
 that can use excess energy from the chemical reactions in  the fuel
(adenosine triphosphate) and perform mechanical work.

It has been understood \cite{TD} that the automaton exhibits self-organized
criticality for the range of values of the control parameter $p\ge p^\star$,
where $p^\star =p_c^{SDP}=0.7054853(5)$ is the percolation threshold for the
site directed percolation  \cite{Jensen}  on the square lattice.
Due to the dynamic conservation
law---conserved number of grains in the interior of the pile---and the
probabilistic  relaxation rules which are locally  like the generalized
site-bond directed percolation (see below), the system
exhibits emergent spatial structure, as discussed in detail in
Ref.\ \cite{TD}.
 Moreover, the avalanche exponents for the
integrated probability distributions of duration $P(T)\sim T^{-(\tau _t-1)}$
and size of avalanches, $D(s)\sim s^{-(\tau_s-1)}$, are expressible in
terms of the standard directed percolation exponents in $d=1+1$ dimensions
as follows \cite{TD}:
\begin{equation}
\tau _t-1\equiv \alpha =(d-1)\zeta _{DP} -\left(\beta /\nu _\|\right)_{DP}
\; ,
\label{exp_relation}
\end{equation}
and $\tau_s=2-1/\tau _t$, and the anisotropy exponent $\zeta =\zeta _{DP}$.
Subscript $DP$ refers to directed percolation, and
$\beta $ and $\nu_\|$ are the critical  exponents for the order parameter
and parallel correlation length, respectively.

In the present work we extend the study of the model of Ref.\ \cite{TD}
in two ways: (1) We study the probability distribution of the
potential energy dissipated in a relaxation event (avalanche)
due to grains drop from higher to lower positions at the fluctuating
sandpile surface. This distribution
is unique for the dynamic sandpile models with ragged spatial structure
and has no  counterpart in the directed percolation processes.
Thus we expect
that the exponents characterizing its scaling properties  are new.
(2) We analyze the behavior of the system close to the
 phase transition point  in terms of $(i)$ scaling properties of the
survival probability distribution for  $p < p^\star$ and $(ii)$
by determining  the time averaged outflow  current $ <J(p)>$, which
behaves as an order parameter. The outflow current results from
avalanches which last longer than the system size  $T\ge L$. The average
is taken over {\it external} time scale, which is measured in number
of added grains. The internal current at time $T<L$ is defined as
 $<j(T,p)> \sim T^{-\alpha }m(T,p)$, where $m(T,p)$ is the average
flux of particles at time $T$, and $T^{-\alpha }$ is the probability
that an avalanche survives $T$ steps.
In the steady state outflow current $ <J(p)>$ balances
the input current, which is one particle per time step, and thus it
is equal to one. For $p<p^\star $ the system ceases to conduct particles,
and the outflow current drops to zero for $\ell \gg \xi (p)$ and finite
lattice size $L$ as
 \begin{equation}
<J(p)> \sim (\delta p)^\beta g(L/\xi (p)) \; ,
\label{OP}
\end{equation}
where $\delta p \equiv (p-p^\star )/p^\star $ measures the distance from
the steady state and $\xi (p) \sim (\delta p)^{-\nu _\| }$ is the parallel
correlation length.
The set of critical exponents is  determined by the appropriate scaling fits
 and using the scaling relations that are valid in
the present  dynamical model (see Sections IV and V).

The organization of the paper is as follows: In Sec.\ II the phase
diagram of the system is calculated numerically for finite lattice size $L$.
In Sec.\ III we study scaling properties of the  distribution of energy
dissipated in avalanches.
In Sec.\ IV we present detailed numerical analysis of the phase transition.
Summary of the universal scaling exponents and the discussion of the
results is given in Sec.\ V.

\section{Phase diagram}

Due to probabilistic dynamic rules the avalanches in this model show a
fractal structure; an example is shown in Fig.\ 1 (bottom). In the
limit $p=1$
this model reduces to the deterministic directed model with compact
avalanches, which has been introduced and solved  exactly in Ref.\ \cite{DR}.

As discussed in detail in Ref.\ \cite{TD}, for  $p^\star \le p < 1$
the relaxation rules {\it at each site} of the system may be visualized
as the  rules of a Domany-Kinzel cellular automaton \cite{DK} of generalized
site-bond directed percolation (DP), with probabilities $P_1$ and $P_2$
that a toppling occurs if one or two particles, respectively, drop
at that site. According to Eq.\ (\ref{rules}), we have $P_1 \equiv p\rho $,
where $\rho $ is the probability that the site has height $h\ge 1$,
 and $P_2 \equiv p$  by definition. In contrast to DP, in the present dynamic
 model the state is being systematically built up after each avalanche,
and the probability $\rho $ was found to vary with the distance   $\ell $
from the top of pile   as \cite{TD}
 $\rho (\ell ,p)= \rho ^\star(p)-A(p)\ell ^{-x}$.
 Here  $x=1/\nu_\|^{DP}$  (see inset to Fig.\ 2)
 is the inverse parallel correlation length exponent of the
 directed percolation \cite{DP-comment}.
 In the above formula  $\rho ^\star (p)$ are
the values of $\rho (\ell ,p)$  reached at $\ell \to \infty$. Notice
that  the distance  $\ell $ from the top of the pile is equivalent to
the duration $T$ of avalanches.
 In Fig.\ 2 we show time-averaged $<\rho ^\star (p)>$ vs. $p$
obtained numerically and averaged over lower third of the lattice
with $L=100$,  for various values of $p$. Two types of initial conditions
are used: ($a$) full lattice (all lattice sites are occupied by at least
 one particle),  and ($b$) half-full lattice (half of the sites which
 are  selected  randomly  have zero heights  and the rest of sites
are occupied).
In both cases, for the initial set of probabilities
$(p,\rho ^\star )$ in the region to the
 right of  $p^\star $ the system self-organizes (after some transient time)
to sitting {\it close} to the DP  critical line (cf. Fig.\ 2).
Left of the line $ p=p^\star$ different initial conditions lead to
separate final states.
We used $8\times 10^6$ time steps for  each point.
 Notice that due to finite size of the lattice
$<\rho ^\star (p^\star)>$ is still somewhat smaller than one (indicated
by dotted line in Fig.\ 2), and that
a dynamical hysteresis occurs in the region $0.5\le p < p^\star $
\cite{comment-dhyst}.
At $p^\star $ an instability---building up of heights---starts at lower
 boundary of the pile and proliferates inside, reaching the first row for
probability of toppling exactly  $p=1/2$. Apart from the finite-size
effects, the phase diagram in Fig.\ 2 is in agreement with general
theoretical considerations given in Ref.\ \cite{TD}.

\section{Dissipated  energy distribution}

Due to the probabilistic character of the dynamic relaxation rules in Eq.\
(\ref{rules}) for $p<1$ and the conservation of particles in the interior
of the pile, our dynamic model exhibits the  emergent spatial
structure \cite{TD}, which is characterized by  rough surface $h(i,j)$
embeded in 3-dimensional space. Therefore,
mass transport in the preferred direction takes place along  a rough
surface. Conditions for the
potential energy  dissipation are fulfilled  when the height difference
$\partial _\|h_\pm \equiv h(i,j)-h(i+1,j_\pm)-1$ along the direction
of transport is positive. More precisely, since relaxation at a site
involves two particles, the  energy is dissipated at that
site when the sum E$(i,j)\equiv \partial _\|h_- +\partial _\|h_+ -1 >0$.

Two comments are in order at this point: (1) Although the driving force
 of the grain transport in this model is {\it not} a gradient of the
potential energy, as discussed in the Introduction, we believe that
cumulative potential energy dissipated in an avalanche is an interesting
quantity which is directly related to temporal fluctuations of the emergent
structure in the real space. We concentrate on the properties of the energy
distribution near the phase transition point $p=p^\star$, where the
surface exhibits dramatic fluctuations, and only briefly discuss scaling
behavior for $p>p^\star$.
(2) We distinguish between energy dissipated at a fixed site, through which
various avalanches run, and the energy dissipated at different points
in the whole avalanche. Here we study scaling properties of the latter
quantity. Notice that the condition  $E(i,j) >0$ is fulfilled at a set of
points ${\cal{S}}$ which is a random subset (see Fig.\ 1) of the avalanche
size $s$ \cite{comment-redsites}. Total energy dissipated in an avalanche is
then $E=\sum _{\cal{S}}E(i,j)$.

The probability distribution of the dissipated energy $P(E)$ is found to
obey a power-law behavior with the exponent $\tau _E $ for
$p^\star\le p <1 $ and  the following  scaling form is satisfied
\begin{equation}
P(E,L) =L^{(\tau _E-1)D_E}{\cal{P}}(EL^{-D_E}) \
\label{pe_fss}
\end{equation}
with $(\tau _E-1)D_E =\tau _t -1$, where $\tau _t-1 \equiv \alpha $ is the
survival probability distribution exponent. In Fig.\ 3 we show the
integrated distribution of dissipated
energies for $p=p^\star$ and for four different values of lattice size $L$.
The slope gives the exponent $\tau _E-1=0.24$,  and
the finite-size scaling plot according to (\ref{pe_fss}), which is
 shown in the inset to Fig.\ 3, is obtained with $\alpha =$0.45 and
 $D_E=$1.84.
In contrast to the survival probability distribution and size distribution
of avalanches, the  distribution of dissipated  energy  cannot be defined
in the directed percolation processes, and thus the exponents $\tau _E$
and $D_E$ are new and  are not directly related to the DP exponents.
Moreover, we find that the exponents $\tau _E$ and $D_E$ are $p$-dependent,
for instance, for $p=$0.8 we obtain  $\tau _E=$ 1.27 and $D_E=$ 1.66,
and  $\tau _E=$ 1.29 and $D_E=$ 1.55 for $p=$0.9.
However, a combination  of these exponents can be related
to the survival probability exponent $\alpha $ {\it via} the  scaling relation
 $(\tau _E-1)D_E =\alpha$, which holds
in the SOC states,  where $\alpha $ is the  universal exponent expressible in
 terms of  DP exponents {\it via} Eq.\ (\ref{exp_relation}). This scaling
relation is satisfied within  numerical  error bars (estimated as  $\pm 0.02$)
 for all values of $p$ in the region $p^\star\le p <1$.
In the limit $p=$1 the critical state is exactly known  \cite{DR}
and consists only of the heights $h=1$ and $h=0$. Consequently,
dissipated energy is bounded to values fife integer values,
and the distribution $P(E)$  has the same scaling exponents as the size of
avalanches distribution.

For $p^\star \le p<1$ emergent spatial structure
appears due to both stochastic dynamics and the conservation of number of
particles in the interior of the pile \cite{TD}. At the edge of the scaling
region ($p=p^\star $) we find that the average height
(averaged in the transverse direction) increases with the distance
$\ell $ from the top row as $<h(\ell )> \approx a\ell ^{b}$,
with $b=0.59\pm 0.02$ (cf. Fig.\ 1 (top)).
Therefore, for large $\ell $ there is a finite probability of large heights,
however, at the same distance $\ell $ some site have height zero, since the
system is in the stationary critical state (in the opposite the
avalanche would
propagate as a directed percolation cluster, which violates stationarity
condition). Thus, $\partial _\| h $ also increases with $\ell $ and becomes
unbounded for $\ell \to \infty $. Thus, the energy cutoff
has additional nontrivial $\ell $-dependence, which is not contained in
the $\ell $-dependence of the  avalanche size cutoff, indicating that
the sandpile surface is a fractal at $p=p^\star $.
In the interior of the scaling region ($p^\star <p<1$), the average
height remains finite and not a function of $\ell $, however, height
distribution does strongly depend on $p$.
We find that the width of the height distribution $w(p)$ increases smoothly
with decreasing $p$ from $w=1$ at $p=1$ to a fast diverging function at
$p\to p^\star$. Neighboring sites are weakly correlated since the dynamics
is governed by the critical height rule only, and thus  dissipated
energy at a site $E(i,j)$ also depends on $p$. We checked by direct
calculation that the distribution of dissipated energy at a fixed site in
the interior of the pile, $P(E_{site})$, taken over
$2\times 10^6$ avalanches exhibits strong $p$-dependence.
It is an exponential function of  width $w_{es}$, which is increasing
smoothly with decreasing $p$ and  becomes nearly power-law at $p=p^\star $.
We believe that the $p$-dependence of the height distribution is the origin
of the observed nonuniversality of the energy exponents.
On the other hand, size and duration of avalanches are governed by the
probabilities $p$ and $\rho ^\star $, which sit always at the DK critical
line in Fig.\ 2.
It should be noticed that the probability $\rho (\ell )$
does not depend on  particular values of  heights $h>1$, and
thus the avalanche exponents remain universal (cf.
Eq.\ (\ref{exp_relation})).

We  find that the distribution of mechanical work done by the system
exhibits a curvature and not a power-law behavior.

\section{Dynamic phase transition}

In the region below $p^\star $ the system ceases to conduct and starts
accumulating  particles. As a consequence the critical steady state is lost
(see detailed discussion in Ref.\ \cite{TD}) and the probability distributions
exhibit exponential cut-offs with  finite correlation length, depending
on the distance from $p^\star $. In  Fig.\ 4 we show the
survival probability distribution $P(T,p)$ for few values of $p < p^\star$
and $L=$200. In general, a distribution
$P(X,p,L)$ satisfies the following scaling form in the subcritical region
\begin{equation}
P(X,p,L)=\left(\delta p\right)^{D_X\nu _\|\tau _X }
{\cal{P}}(X\left(\delta p\right)^{D_X\nu _\|}, XL^{-D_X}) \  ,
\label{p-scal}
\end{equation}
where $\delta p \equiv (p^\star -p)/p^\star $ and $X$ stand for $T$,
$s$, or $E$, respectively, and $D_X$ is the  corresponding
fractal dimension. In the case of distribution of durations $P(T,p,L)$
we have $D_T\equiv z$ the dynamic exponent, and  $z=1$
in the present model. Therefore there are no finite-size effects in the
survival probability distribution, which makes it particularly suitable
for the subcritical scaling analysis. In the case of size  and energy
 distributions, one is restricted to values of $p$ and $L$
such that the condition $(\delta p)^{-\nu _\|}/L \ll 1$ is satisfied.
In the inset to  Fig.\ 4 the scaling collapse according to Eq.\ (\ref{p-scal})
of the survival probability distribution is shown, where we have used
$\alpha =0.45$  and $z\nu _\| =1.22$.

Another way to study the dynamic phase transition is by direct
measurements of the order parameter, i.e., the time-averaged outflow
current $<J(p)>$.
In the critical steady state $<J(p)>=1$, thus  balancing the average
input current. Below the transition point this balance is lost.
The outflow current decreases
reaching zero at some lower value of $p$, which depends
 on the system size $L$ [see Fig.\ 5 (bottom)].
For different lattice sizes we expect the following scaling form to hold:
\begin{equation}
<J(p,L)> = L^{-\beta /\nu _\|}{\cal{J}}(L^{1/\nu_\|}(p-p^\star)/p^\star) \  .
\label{op-fss}
\end{equation}
This scaling form follows from general scaling properties of the {\it internal}
current for $T<L$, that reads:
$<j(T,p,L)> \sim  L^{\lambda _J}j(L^{1/\nu_\|}(p-p^\star)/p^\star,L^{-z}T)$.
By choosing $L\sim (p/p^\star -1)^{-\nu_\|} \equiv \xi $ and having  defined
the exponent $\beta $ in  Eq.\ (\ref{OP}), we find  that the anomalous
dimension $\lambda _J=-\beta/\nu_\|$. Therefore
$<j(T,p,L)> = L^{-\beta /\nu _\|}{\cal{G}}(L^{1/\nu_\|}(p-p^\star)/p^\star,
L^{-z}T)$.
In the stationary state for $p \ge p^\star$, correlation length
$\xi \to \infty$ and  the first argument in ${\cal{G}}$ can be neglected.
 For $T\ll L^z$ we expect that the scaling
 function ${\cal{G}}$ behaves as a power, i.e., $<j(T,p,L)>
\sim const \times T^{-\alpha }L^{z\alpha -\beta/\nu_\|}$, which
should be independent on $L$, thus  leading to $\beta /\nu_\| = z\alpha $.
For the {\it outflow} current, however,  we have $T\ge L$ and second argument
of ${\cal{G}}$ can be neglected. Then for $p <p^\star $ one gets the
expression (\ref{op-fss}). Taking $L^{1/\nu_\|}\sim (p-p^\star)/p^\star$
leads to Eq.\ (\ref{OP}).

The scaling plot  according to Eq.\ (\ref{op-fss})
is shown in Fig.\ 5 (top) where we have $\beta/\nu_\| =0.45$
and $1/\nu _\| = 0.83$. Together with the above results and observing
the error bars for $\alpha $ from \cite{TD}, we estimate the
exponents as $\nu _\| =1.22\pm 0.02$ and $\beta =0.56 \pm 0.02$.

Notice that the above values of the  exponents $\nu_\|$ and $\beta $
are close to the values $1.28\pm 0.06$ and
$0.58\pm 0.06$, respectively, obtained by
Monte Carlo simulations  for  $d=3$ dimensional directed percolation
in Ref.\ \cite{Grassberger}. The reason for this similarity lies
in the altered character of the dynamics of our model  below the transition.
Namely, the probability  $\rho ^\star $ of having  height $h\ge 1$ reaches
unity (in the limit of large $L$) at $p^\star $.
The consequences of this are twofold:
($i$) the {\it threshold character} of the
dynamics is lost for $p<p^\star $, i.e., each  site in the lattice
 satisfies  the threshold  condition $h\ge h_c=2$ when a single particle
drops on that site;
 ($ii$) since $\rho =1$ we have
$P1=P2=p$, thus the system spreads the perturbation in the
$(R_\|,R _\bot)$  plane  with the
probability $p$ and builds up heights with probability $q=1-p$.
Therefore, for $0.5<p<p^\star $ we have a dynamic model in which an
avalanche propagates effectively as a cluster in a 3-dimensional
directed percolation with finite widths $\xi_\|, \xi _\bot $
in the plane, and percolating in the vertical direction. However,
there are considerable differences between these processes and
 conventional 3-dimensional DP, leading to generally different
set of exponents, as discussed below.
 An added particle moves along rough surface
$h(R_\|,R _\bot)$, which fluctuates inside the correlated region
$(\xi_\|, \xi _\bot) $. The internal time scale becomes bounded by finite
$\xi _\|$, and the system percolates for $t\to \infty$, where $t$
is now the external time scale (measured by the number of added particles).
Flights of particles along the rough surface are proportional to
local height gradients $\partial _\| h$, which are usually larger than one,
in contrast to contact percolation processes. However, the average
height of the pile $<h>$ grows exactly by one unit with each added
 particle  as a consequence of the conservation of number of particles.

\section{Discussion and Conclusions}

The dynamic model with stochastic relaxation rules of Eq.\ (\ref{rules})
exhibits the universal self-organized criticality for the range of
toppling probabilities $p\le p^\star <1$, and the dynamic  phase transition
at $p=p^\star $. As discussed in detail in Ref.\ \cite{TD}, the avalanche
exponents $\alpha$ for survival probability, $\tau \equiv \tau _s-1 $
for integrated cluster size distribution, and $\zeta $ for the average
transverse extent of clusters, are
expressible in term of standard directed percolation exponents in all
dimensions {\it via} Eq.\ (\ref{exp_relation}).
Here we have shown that the dissipated energy distribution,
which is peculiar to   the dynamic model and has no  analogue in the
directed percolation processes, is
described  by a new exponent $\tau _E$ and corresponding
fractal dimension $D_E$. These exponents appear to be $p$-dependent, however,
their product is related to the universal survival probability exponent
due to scaling relation $(\tau _E-1)D_E =\alpha$.
Power-law behavior of the distribution
$D(E)$ at $p=p^\star $ indicates that the sandpile surface
$h(i,j)$ is a fractal.
We estimate the roughness exponent $\chi$ by measuring the time averaged
height  vs. transverse dimension of the pile $<h(R_{tr})> \sim R_{tr}^\chi $
at various distances $\ell $ from the first  row [see Fig.\ 1 (top)].
By box counting we find that the contour curve  of the
perpendicular section through the pile
 for $p=p^\star$ has fractal dimension
$d_f=1.44\pm 0.045$, leading to $\chi = d_f-1 =0.44\pm 0.04$.
Error bars are estimated from several separate measurements
at different sections. The roughness exponent appears to be
 larger than the one measured in ricepile model
with critical slope rules, where it was found $\chi _{RP}=$0.23
\cite{Al_tracer,Maya}.
In the steady state the height  fluctuates around the  average value
$<h> =19 \pm 6$, increasing
 by one unit at boundary sites of an avalanche and decreasing
by one unit at transport sites with one preceding active neighbor.
 The flame-like profile in Fig.\ 1 (top) indicates individual
site fluctuations, in agreement with the critical height rules in our model.
Closeness of the exponents $\chi \approx \alpha $ indicates that the
number of sites at which the pile grows  is on the average  equal to the
 number of transport sites, i.e.,  the  avalanches have almost no compact
parts (cf. Fig.\ 1).
Below the transition point the pile grows indefinitly for $t\to \infty $,
as discussed in Sec.\ IV.

The dynamic phase transition at $p^\star$ is characterized by the
exponents of the order parameter $\beta $, and the parallel correlation
length  $\nu_\| $, the  numerical values of which appear to be very close to
those of the  directed percolation in $d+1$ dimensions.
However, the exponent $\gamma $ for the order-parameter
fluctuations and the exponent  $\kappa $, which describes the average cluster
growth at time  $T$ as  $m(T)\sim T^\kappa $,  appear to be
different from $d=3$ DP exponents (see Ref.\ \cite{Grassberger}).
A complete set of exponents is given in Table\ I.

With  regard to the exponents in Table\ I
we would like to point out the following: (1)
In the {\it critical steady} state the average cluster growth balances
the average input of particles, i.e.,  $m(T)\sim 1$, leading to
 $\kappa =0$. The
 scaling relation  $\kappa =\gamma/\nu_\| -1 =D_\|-1-\alpha $ holds,
thus we have  $\gamma =\nu _\|$;
(2) At the dynamic phase transition the hyperscaling relation (HS)
$2\beta +\gamma =[1+(d-1)\zeta]\nu_\| $ appears to be violated
\cite{HS-comment,DickmanTrety},
in contrast to the thermodynamic DP phase transition (cf. Table\ I).
The avalanche exponents are determined in  Ref.\ \cite{TD}.
The exponents for DP in $d=2$ are taken from Ref.\ \cite{Jensen} and
the corresponding avalanche exponents $\tau _{DP}$ and $D_{DP}$
are calculated using the scaling relations.
(Notice that due to the presence of anisotropy, at least three exponents
should be known in order to determine completely the universality class.)
The following scaling relations are valid both in DP and in dynamic
models: $\beta /\nu_\| =\tau _t-1\equiv \alpha$;
$\beta /(\beta +\gamma ) =\tau _s -1\equiv \tau $;
$\zeta =\nu_\bot /\nu_\| $, and
$D_\|\nu_\|=\beta +\gamma $.
The hyperscaling violation exponent $\Omega $ is defined via
$\beta +\gamma =[1+(d-1)\zeta -\beta/\nu_\| +\Omega ]\nu_\|$, which
together with the above equations leads to
$\Omega =D_\| -1+\alpha -(d-1)\zeta $. Here  $D_\|$ is the fractal dimension
of the size of relaxation clusters measured
with respect to the length parallel to the transport direction.
Using the fact that $D_\|=1+\alpha $ in the dynamic model,
we may write $\Omega = 2\alpha - (d-1)\zeta$.

In the directed dynamic processes it is
useful to define another  exponent $\beta ^\prime$, such that the
generalized hyperscaling relation \cite{DickmanTrety,multi-absorbing,BARW}
\begin{equation}
\alpha (1 + \beta ^\prime /\beta ) +\kappa = (d-1)\zeta \; ,
\label{gen-HS}
\end{equation}
is satisfied. Here $\beta ^\prime $ is related to the ultimate survival
 probability (the survival probability of a cluster grown from a fixed
seed), whereas $\beta $ governs the usual
order-parameter---stationary density of active sites.
Recently exponential inequality $\beta ^\prime \neq \beta $ was found in
 models with multiple absorbing configurations \cite{multi-absorbing}
and in branching annihilating random walks with even parity \cite{BARW}.
 Our results in this paper suggest
that in the self-organized dynamic critical states in directed models
$\kappa \equiv 0$ and  $\beta \neq \beta ^\prime$ is
always satisfied.  We have
\begin{equation}
\Omega = {{\beta }\over{\nu _\|}} -{{\beta ^\prime }\over{\nu _\|}} >0 \; .
\label{BB}
\end{equation}
Using the above scaling relations and Eq.\ (\ref{exp_relation}) we find that
$\beta ^\prime /\nu _\| = \alpha _{DP} $=0.159 (cf. Table\ I). Therefore,
the HS violation exponent $\Omega $ is given by the difference between the
survival probability exponents in the dynamic model and the underlying
directed percolation as $\Omega = \alpha - \alpha _{DP}$, which turns to be
equal to the cluster growth exponent of DP, i.e.,  $\Omega =\kappa _{DP}$.
One can also define a new configuration exponent $\gamma ^\prime $ via
$\gamma ^\prime /\nu _\| \equiv \gamma /\nu _\| + \Omega $, such that
the relation $2\beta ^\prime + \gamma ^\prime = (1 +(d-1)\zeta )\nu _\| $
is satisfied. By inserting the above expression for $\Omega $ into
$\gamma ^\prime  $, we have  $\gamma ^\prime /\nu _\|= 1+\kappa _{DP} =
1.314$. Using value of $\nu _\|$ from Table\ 1 we find  $\gamma ^\prime
=1.586\pm 0.026 $, and $\beta ^\prime =0.195\pm 0.012 $.
The origin of the difference between $\alpha$ and $\alpha _{DP}$ lies
in the dynamic conservation law, which leads to the emergent spatial
structure and dependence of the branching probability $\rho (\ell )$ on
distance $\ell $, as discussed in Ref.\ \cite{TD}. Our present
results suggest that the dynamic conservation law is also responsible for
the new universality class of the dynamic phase transition at the edge of
the critical region.

\acknowledgments
I thank Deepak Dhar for fruitful discussions and suggestions which
led to the results presented in Fig.\ 2.  I also thank Maya Pazsuski
for helpful comments and suggestions.
This work was  supported by the Ministry
of Science and Technology of the Republic of Slovenia.

\narrowtext
 \begin{table}[ht]
 \label{Table1}
 {\caption{Critical exponents in stochastic directed model (PH)
 and in directed percolation (DP) in $d=2$ dimensions. Also shown are
exact exponents in $p=1$ limit (DR). }}
  \begin{center}
\begin{tabular}{|c||c|c|c|c|c||c|c|c|c|}
 M-E&$\alpha $&$\tau  $& $D_\|$& $\zeta $ &$\kappa $ &$\beta $&$\nu _\| $
 &$\gamma $&$\Omega $\\
\hline\hline
PH & 0.46 & 0.31  &  1.46  & 0.62& 0& 0.56 &1.22 &1.22&0.30 \\
\hline
 DR   & 1/2 & 1/3 & 3/2 & 1/2& 0& -& -& -&1/2\\
\hline
  DP  &  0.159&  0.108&  1.472& 0.634&0.314& 0.276& 1.734& 2.278&0\\
\end{tabular}
\end{center}

\narrowtext
\begin{figure}[thb]
\epsfxsize=82mm\epsffile[54 68 523 578]{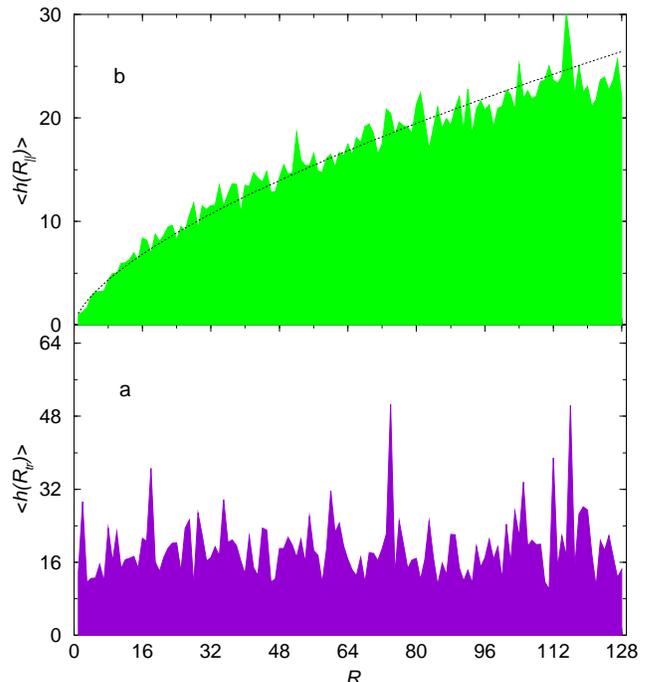}
\epsfxsize=82mm\epsffile[41 244 568 547]{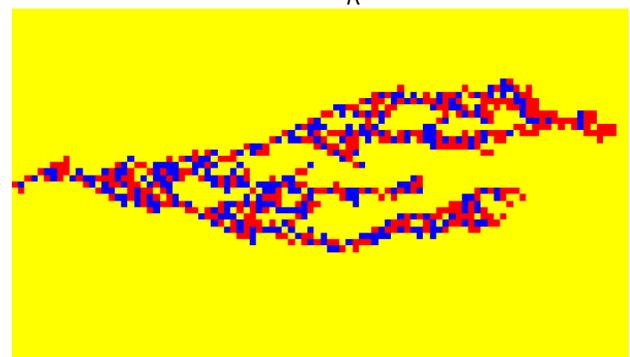}
\caption{\label{fig1} Bottom: Avalanche in $1+1$ stochastic
  model for $p=p^\star $ and $L=$128. Transport direction is
from left to right. Dark points represent sites  at which system
performs work, and at light dark points it dissipates  energy.
Top: (a) Transverse section of
the pile at distance $\ell =96$ from the top. Heights are
averaged over $5\times 10^6$ runs. (b) Height (averaged in transferse
direction) vs. parallel distance from the top in the stationary state.
 Dotted line: $1.6\ell ^{0.59}$. Heights are measured in number of
grains and  $R$ is given in lattice sites.}
\end{figure}

\begin{figure}[thb]
\epsfxsize=82mm\epsffile[37 67 563 563]{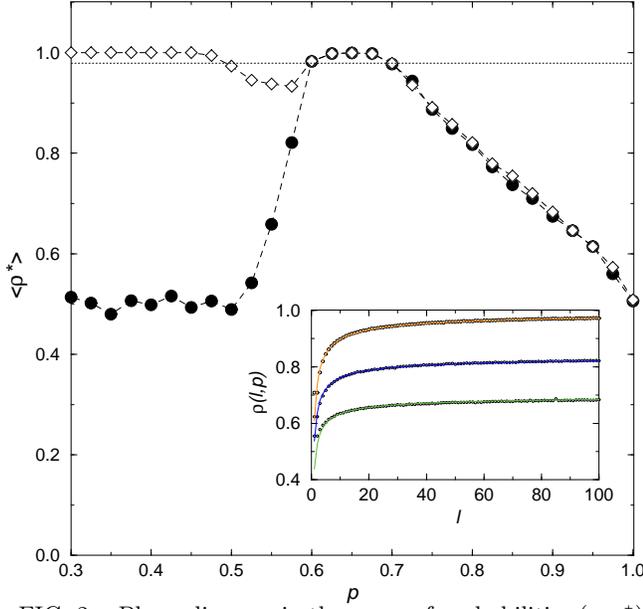}
\caption{\label{fig2} Phase diagram in the space of probabilities
 $(p,\rho ^\star )$ obtained numerically for a lattice with $L=100$.
 Open symbols correspond to initial conditions along the line $(p,1)$,
and filled symbols to  initial conditions along the
line $(p,1/2)$. Dotted line indicates value $<\rho ^\star>$ at $p^\star$.
Inset: $\rho (\ell)$ vs. $\ell $ for (top to bottom) $p=p^\star $, 0.8,
and 0.9, with fit lines described in the text.}
\end{figure}

\begin{figure}
\epsfxsize=82mm\epsffile[37 67 563 563]{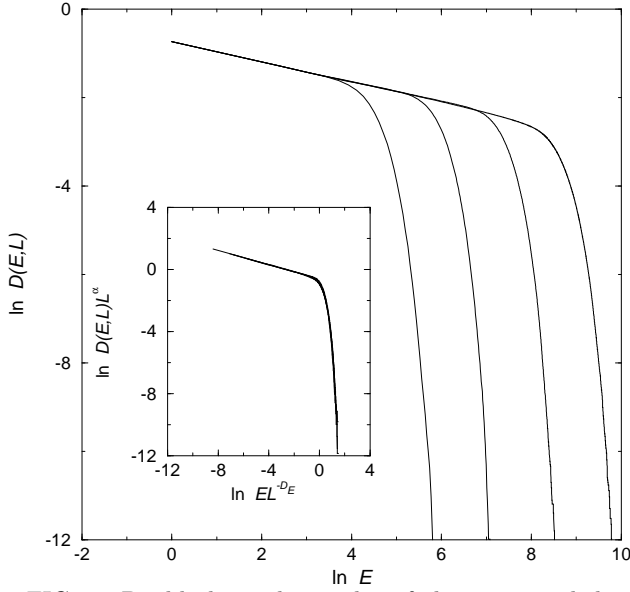}
\caption{\label{fig3}Double logarithmic plot of the integrated
distribution of dissipated energies $D(E,L)$ vs.  $E$ for $p=p^\star $
and for  $L=12,24,48,$ and $96$ (left to right).
 Inset: Scaling plot according to
Eq.\ (\ref{pe_fss}) with $\alpha =0.45$ and $D_E=1.84$. }
\end{figure}

\begin{figure}
\epsfxsize=82mm\epsffile[37 67 563 563]{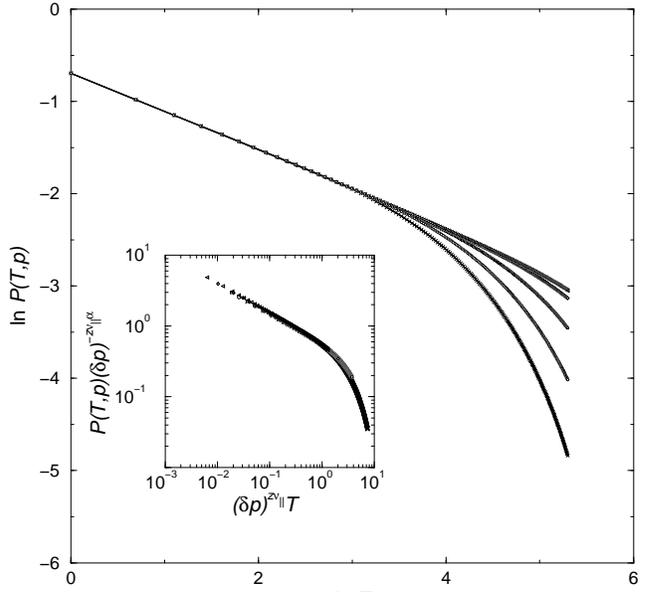}
\caption{\label{fig4}Double logarithmic plot of the integrated
distribution of durations $P(T,p)$ vs.  $T$ for $L$=200,
and $p$=0.695, 0.69, 0.68, 0.67, and 0.66 (top to bottom)
in the subcritical region. Inset: Scaling plot according to
Eq.\ (\ref{p-scal}) with $\delta p \equiv (p^\star -p)/p^\star $
and exponents $\alpha =0.45$ and $z\nu _\|=1.22$. }
\end{figure}

\begin{figure}
\epsfxsize=82mm\epsffile[28 68 507 751]{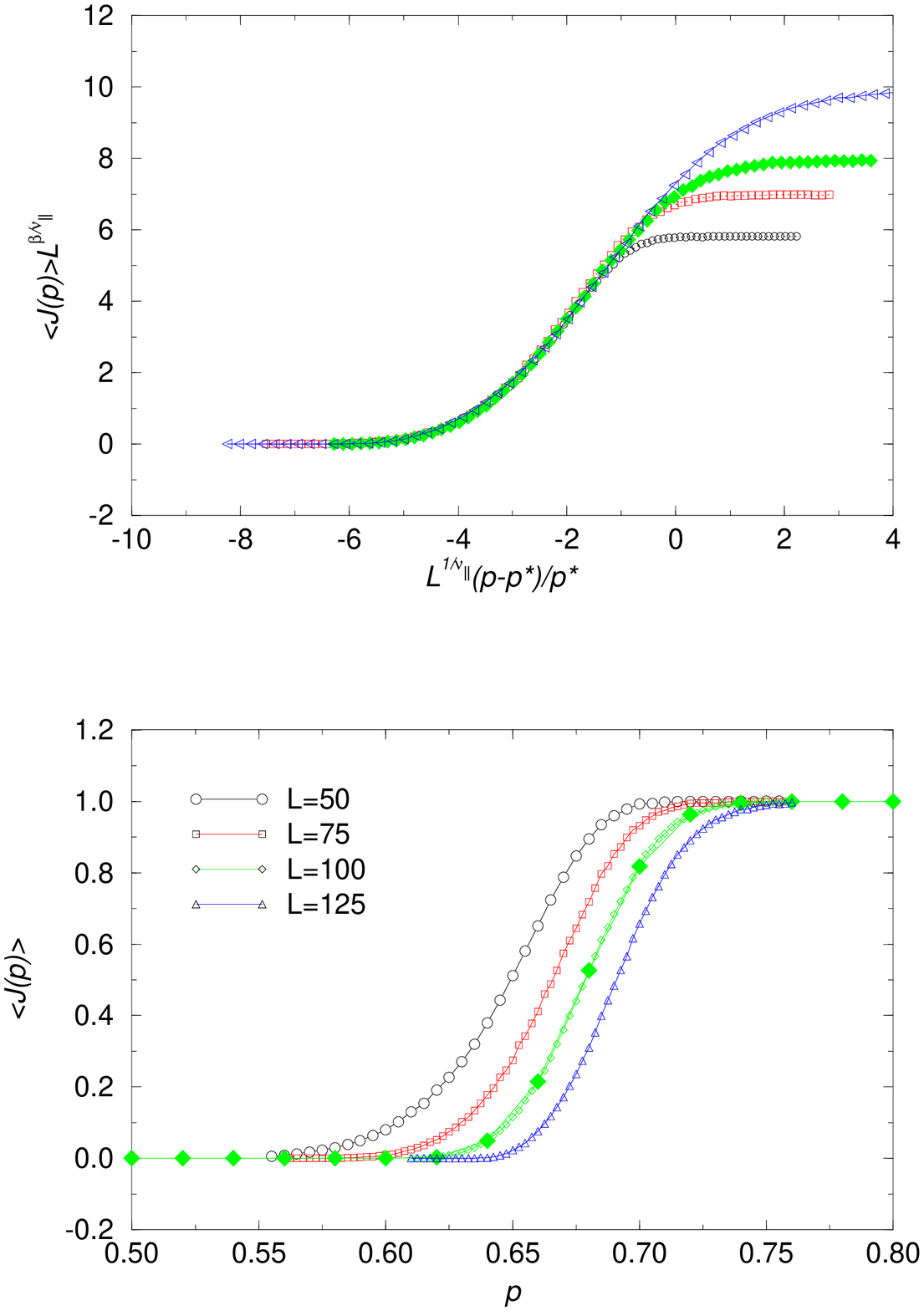}
\caption{\label{fig5}Average outflow current $<J(p)>$ vs. $p$
for various lattice sizes (bottom). Finite size scaling plot according to
Eq.\ (\ref{op-fss}) with $\beta/\nu_\|=0.45$ and $1/\nu_\|=0.83$ (top). }
\end{figure}

\end{table}

\end{multicols}


\begin{references}
\bibitem{DP-comment} For $d=2$ the exponents of directed percolation
are known very accurately, see Ref.\ \cite{Jensen}.
\bibitem{comment-dhyst}Notice that $<\rho^\star >$ can be viewed as
a response function of the system to a slowly increasing external ``field''
by one unit (particle) per time step.
\bibitem{comment-redsites}Set of red sites is a fractal with the same
scaling properties  as avalanche size.
\end{references}
\end{document}